\documentstyle[12pt]{article}

\newlength{\extraspace}
\newlength{\extraspaces}
\catcode`\@=11
 
\def\numberbysection{\@addtoreset{equation}{section}
\def\theequation{\arabic{section}.\arabic{equation}}}

\begin{document}
%
%\addtolength{\baselineskip}{.7mm}
\thispagestyle{empty}
\begin{center}
\begin{flushright}
TIT/HEP--358 \\
{\tt hep-th/9701155} \\
January, 1997 \\
\end{flushright}
\vspace{3mm}
\begin{center}
{\Large
{\bf Duality in Supersymmetric $SU(N)$  Gauge Theory with a Symmetric Tensor 
}} 
\\[18mm]

{\sc Tadakatsu Sakai}\footnote{
\tt e-mail: tsakai@th.phys.titech.ac.jp} \\[3mm]
{\it Department of Physics, Tokyo Institute of Technology \\[2mm]
Oh-okayama, Meguro, Tokyo 152, Japan} \\[4mm]

\end{center}
\vspace{18mm}
{\bf Abstract}\\[5mm]
{\parbox{13cm}{\hspace{5mm}
Duality in supersymmetric $SU(N)$ gauge theory with a symmetric tensor 
is studied using the technique of deconfining and Seiberg's duality. 
By construction the gauge group of the dual theory necessarily becomes a 
product group. In order to check the duality, several nontrivial consistency 
conditions are examined. In particular we find that by deforming along a flat 
direction, the duality flows to the Seiberg's duality of $SO(N)$ gauge theory.

PACS numbers: $11.15.{\rm -q}, 11.15.{\rm Tk}$, $11.30.{\rm Pb}$

Key words: supersymmetric gauge theory, duality
}}
\end{center}
\vfill
\newpage
\vfill
\newpage
%\addtocounter{section}{1}
%\setcounter{equation}{0}
%\setcounter{section}{0}
%\setcounter{footnote}{0}
%\numberbysection

%%%%%%%%%%%%%%%%%%%%%%%%%%%%%%%%%%%%%%%%%%%%%%%%%%%%%%%%%%%%%
%%%%%%%%%%%%%%%%%%%%%%%%%%%%%%%%%%%%%%%%%%%%%%%%%%%%%%%%%%%%%
%%%%%%%%%%%%%%%%%%%%%%%%%%%%%%%%%%%%%%%%%%%%%%%%%%%%%%%%%%%%%

%\documentstyle[12pt]{article}
%\hyphenation{Cou-lomb}
%\begin{document}

%\begin{flushright}
%6th Edition \\
%\today
%\end{flushrigh}

%\vspace{7mm}
%\pagebreak[3]
%\addtocounter{section}{1}
%\setcounter{equation}{0}
%\setcounter{subsection}{0}
%\setcounter{footnote}{0}
%\begin{center}
%{\large {\bf \thesection. Introduction }}
%\end{center}
%\nopagebreak
%\medskip
%\nopagebreak
%\hspace{3mm}

Recent progress in supersymmetric gauge theories made it possible to argue 
non-perturbative aspects of them \cite{Sei1} \cite{IRSei} (for a review, see 
\cite{IS;lec}). 
In particular, Seiberg has shown \cite{Sei2} that SUSY gauge theories have a 
dual description in terms of a SUSY gauge theory with a different gauge group. 
For example, $SU(N_c)$ gauge theory with $N_f$ 
flavors is dual to $SU(N_f-N_c)$ gauge theory with $N_f$ flavors. 
Although both theories have different gauge groups, global 
symmetries of the two theories are the same and they satisfy several 
nontrivial consistency checks on the duality. 
Duality in SUSY gauge theories is also studied in 
\cite{KSS} \cite{IS;so} \cite{IP;sp} \cite{IRSt}
\cite{Chi-nonChi} \cite{Decon} \cite{Prod} \cite{Excep} \cite{Aha} 
\cite{RS} \cite{Maru}. 

Duality in SUSY gauge theories with tensor matters is 
extensively investigated. 
One way to do it is to use the technique of deconfining \cite{Decon}. 
That is, a tensor 
matter is considered to be a bound state of a confining gauge theory. 
The dual description of an original theory can be obtained by dualizing 
the deconfined theory instead of the original theory. One 
of the merits of the deconfining technique is that one can construct dual 
theories by using Seiberg's duality. This is because all the matter fields of 
the deconfined theory belong only to the fundamental representation of 
the gauge 
group. The gauge group of the deconfined theory is necessarily a product 
group, and thus the dual theory 
also has a product gauge group, even though the original theory has a 
single gauge group. 

In this letter we consider duality in 
$SU(N)$ gauge theory with a symmetric matter. 
We first discuss the non-perturbative superpotential and 
the quantum moduli space. Using 
holomorphy and symmetry, it turns out that no superpotential is dynamically 
generated, and at the origin of the moduli space the theory is in 
non-abelian Coulomb phase. 
This indicates that the $SU(N)$ gauge theory with a symmetric tensor has a 
dual description. 
Next we construct the deconfined theory. It is argued that the original 
theory can be deconfined by introducing an $SO(N+5)$ gauge group. Then we use 
Seiberg's duality to obtain the dual theory. Note that by virtue of 
Seiberg's duality, 
global symmetries of the dual theory are the same as those of the 
original theory, 
and the 't Hooft anomaly matching conditions are trivially satisfied. 
The mappings of the gauge invariant operators between the two theories 
are easy to establish. 
It is also argued that by dualizing the dual theory again, we can obtain 
another dual theory which has different matter contents while having the 
same gauge group. 
As a consistency check on the duality, we turn on a vacuum expectation value 
in the original theory. We find that the dual theory is properly deformed 
as the vev's are turned on in the original theory. In particular we show that the 
duality in $SU(N)$ gauge theory with a symmetric matter flows to the 
Seiberg's duality in $SO(N_c)$ gauge theory by deforming it along a 
flat direction.

We start with the theory; 
\begin{equation}
\label{electric;theory}
\matrix{   
&SU(N)&SU(F)&SU(N+F+4)&U(1)_Q& U(1)_{\bar{Q}}&U(1)_S &U(1)_X  \cr
Q^{xA}           &\Box&\Box&{\bf 1}&1&0&0&0  \cr
\bar{Q}^l_A     &~\Box^{\ast}&{\bf 1}&\Box&0&1&0&0  \cr
S^{AB}           &\Box\;\!\!\!\Box&{\bf 1}&{\bf 1}&0&0&1&0  \cr}
\end{equation}
where $A,B=1,2,\cdots, N,~~ x=1,2,\cdots,F$ and 
$l=1,2,\cdots, N+F+4$. 
We refer to this as the electric theory. 
All the global $U(1)$'s are anomalous, and can be used as selection 
rules by assigning appropriate charges 
to the dynamically generated scale $\Lambda_{SU(N)}$ 
\begin{equation}
\label{selectionrule}
\matrix{&U(1)_Q& U(1)_{\bar{Q}}&U(1)_S &U(1)_X  \cr
\Lambda_{SU(N)}^b &F&N+F+4&N+2&-2(F+3) \cr}
\end{equation} 
where $b=2N-F-3$ is the coefficient of the $1$-loop beta function.

From the four anomalous $U(1)$'s, three anomaly free $U(1)$'s can be 
constructed. One of them is $R$-symmetry. Here we take the following 
charge assignments 
\begin{equation}
\label{anomalyfree;u1}
\matrix{&U(1)_1& U(1)_2&U(1)_R  \cr
Q &1&N-F&{2(F+3) \over F} \cr 
\bar{Q} &{-F \over N+F+4}&F&0 \cr
S &0&-2F&0 \cr}
\end{equation} 

Using holomorphy and symmetry, it is easy to see that no superpotential can 
be generated for all $F$. Thus the flat directions are left 
quantum mechanically. It is useful to describe the flat directions in terms of 
gauge invariant operators, some of which are given by 
\begin{eqnarray}
&&M = Q\bar{Q}, \quad H = S\bar{Q}\bar{Q}, \quad U = \det S, \nonumber \\
&&\bar{B} = \epsilon~\bar{Q}^N, \quad
V_k = \epsilon~Q^k(S\bar{Q})^{N-k}, ~~{\rm for}~k \le F, 
\label{gaugeinvariants}
\end{eqnarray}
where $\epsilon$ is the epsilon tensor of the $SU(N)$. 
All the gauge invariant operators are not independent and satisfy 
some constraint equations. 
Note that the classical constraints hold quantum mechanically, 
because the $U(1)_X$ charge of the dynamical scale $\Lambda_{SU(N)}$ 
is always nonzero. 
Thus the quantum moduli space is the same as the classical 
one, so that the singularity at the origin of the classical moduli space 
cannot be smoothed out. In order to discuss the physical meaning of the 
singularity, consider the 't Hooft anomaly matching conditions. 
It turns out that the singularity cannot be attributed to the appearance of 
singlet fields. 
This implies that our model is in 
non abelian Coulomb phase for $F \le 2N-3$, which is expected to have a dual 
description.  

As discussed above, our strategy to find the duality is to construct the 
deconfined theory of (\ref{electric;theory}) by using the technique of 
deconfining \cite{Decon}. 
We first consider $SO(N+5)$ SUSY 
gauge theory with $N+1$ fundamentals $y^A$ and $z$, and $N+1$ singlets 
$\bar{P}_A$ and $u$, where $A=1, 2,\cdots, N$. Here we assume 
this theory has the superpotential at the tree level 
\begin{equation}
W=yz\bar{P}+z^2u.
\label{superpotential;expanded}
\end{equation}
According to the result of \cite{IS;so}, there is a branch of the moduli space 
in which no superpotential is generated, confinement occurs and the massless 
fields at the origin of the branch are given by the mesons 
\begin{equation}
S^{AB}=y^Ay^B,~~P^A=y^Az,~~R=z^2.
\label{meson;expanded}
\end{equation}
The superpotential (\ref{superpotential;expanded}), however, 
makes massive the 
singlet fields $P^A, R, \bar{P}_A$ and $u$, so that the massless field is 
only $S^{AB}$. Now gauge the global $SU(N)$ symmetry weakly. This means that 
the dynamical scales of the two gauge groups satisfy 
$\Lambda_{SU(N)} \ll \Lambda_{SO(N+5)}$. In order for the $SU(N)$ 
to be anomaly free, we must introduce $F$ fundamentals $Q^{xA}$, and 
$N+F+4$ anti-fundamentals $\bar{Q}_A^l$. This is indeed the electric theory 
(\ref{electric;theory}). 
We refer to it as the expanded theory. It follows from 
(\ref{superpotential;expanded}) and (\ref{meson;expanded}) that 
the $U(1)$ charges of the fields are fixed as follows 
\begin{equation}
\label{expanded;theory}
\matrix{ &SU(N)&SO(N+5)&SU(F)&SU(N+F+4)&U(1)_1&U(1)_2&U(1)_R \cr 
y_i^A    &\Box&\Box&{\bf 1}&{\bf 1}&0&-F&0 \cr 
z_i        &{\bf 1}&\Box&{\bf 1}&{\bf 1}&0&NF&-2 \cr 
\bar{P}_A &~\Box^{\ast}&{\bf 1}&{\bf 1}&{\bf 1}&0&-F(N-1)&4 \cr 
u        &{\bf 1}&{\bf 1}&{\bf 1}&{\bf 1}&0&-2NF&6 \cr 
Q^{xA}   &\Box&{\bf 1}&\Box&{\bf 1}&1&N-F&{2(F+3) \over F} \cr
\bar{Q}_A^l &~\Box^{\ast}&{\bf 1}&{\bf 1}&\Box&{-F \over N+F+4}&F&0 }
\end{equation}

Let us now derive the dual description of the electric theory 
from the expanded theory. For convenience, we shall consider the case 
$\Lambda_{SO(N+5)} \ll \Lambda_{SU(N)}$. 
%Now dualize $SU(N)$ using the Seiberg's duality [Seib]. To apply it to 
%the expanded theory, one must work  
%in the case $\Lambda_{SO(N+5)} \ll \Lambda_{SU(N)}$. 
At the scale $\Lambda_{SU(N)}$, 
the expanded theory can be considered to be $SU(N)$ SUSY gauge theory 
with $N+F+5$ flavors. Note that this theory allows a dual description for 
all $F$. Following \cite{Sei2}, consider the meson fields 
\begin{eqnarray}
&&M^{xl}=Q^{xA}\bar{Q}_A^l, \quad N^x=Q^{xA}\bar{P}_A, \nonumber \\
&&C^{li}=y^{iA}\bar{Q}_A^l, \quad D^i=y^{iA}\bar{P}_A.
\end{eqnarray}
Dual quarks which couple to the mesons are also introduced. Note that 
the superpotential (\ref{superpotential;expanded}) makes $D^i$ and $z$ 
massive. Integrating out these fields, we obtain the dual theory: 
\begin{equation}
\label{dual1;theory}
\matrix{ 
&SU(F+5)&SO(N+5)&SU(F)&SU(N+F+4)&U(1)_1&U(1)_2&U(1)_R  \cr 
\bar{q}_{xA} & ~\Box^{\ast}&{\bf 1}& ~\Box^{\ast}&{\bf 1}&{-5 \over F+5}
&-N&-{10(F+3) \over F(F+5)} \cr 
\bar{x}_{iA} &~\Box^{\ast}&\Box&{\bf 1}&{\bf 1}&{F \over F+5}&0
&{2(F+3) \over F+5} \cr 
q_l^A &\Box&{\bf 1}&{\bf 1}&~\Box^{\ast}&{-F(N-1) \over (F+5)(N+F+4)}&
0&{4 \over F+5} \cr 
p^A &\Box&{\bf 1}&{\bf 1}&{\bf 1}&{-F \over F+5}&NF
&-{4(F+4) \over F+5} \cr 
u &{\bf 1}&{\bf 1}&{\bf 1}&{\bf 1}&0&-2NF&6 \cr 
M^{xl} &{\bf 1}&{\bf 1}&\Box&\Box&{N+4 \over N+F+4}&N&{2(F+3) \over F} \cr 
N^x &{\bf 1}&{\bf 1}&\Box&{\bf 1}&1&-N(F-1)&{6(F+1) \over F} \cr 
C^{li} &{\bf 1}&\Box&{\bf 1}&\Box&{-F \over N+F+4}&0&0 \cr }
\end{equation}
The superpotential is given by 
\begin{equation}
W=Mq\bar{q}+Np\bar{q}+Cq\bar{x}+(p\bar{x})^2u.
\label{superpotential;dual1}
\end{equation}
We refer to this theory as the dual I theory. Note that the 't Hooft anomaly 
matching conditions are satisfied.

At the scale $\Lambda_{SO(N+5)}$, the dual I theory can be considered as 
$SO(N+5)$ gauge theory with $N+2F+9$ flavors, because the $SU(F+5)$ gauge 
coupling is much weaker than that of the $SO(N+5)$. 
Thus we can dualize it by using the duality of Seiberg. 
Following the procedures of \cite{Sei2}, consider the meson 
fields 
\begin{equation}
\bar{S}_{AB}=\bar{x}_{iA}\bar{x}_{iB}, \quad H^{lm}=C^{il}C^{im}, \quad 
I^l_A=\bar{x}_{iA}C^{li}. 
\end{equation}
We must also introduce dual quarks which couple to the mesons. Note that 
$q$ and $I$ become massive due to the superpotential 
(\ref{superpotential;dual1}). Integrating them out, the dual theory is 
\begin{equation}
\label{dual2;theory}
\matrix{ &SU(F+5)&SO(2F+8)&SU(F)&SU(N+F+4)&U(1)_1&U(1)_2&U(1)_R \cr 
\bar{q}_{xA} & ~\Box^{\ast}&{\bf 1}& ~\Box^{\ast}&{\bf 1}&{-5 \over F+5}
&-N&-{10(F+3) \over F(F+5)} \cr 
p^A &\Box&{\bf 1}&{\bf 1}&{\bf 1}&{-F \over F+5}&NF&-{4(F+4) \over F+5} \cr 
u &{\bf 1}&{\bf 1}&{\bf 1}&{\bf 1}&0&-2NF&6 \cr 
M^{xl} &{\bf 1}&{\bf 1}&\Box&\Box&{N+4 \over N+F+4}&N&{2(F+3) \over F} \cr 
N^x &{\bf 1}&{\bf 1}&\Box&{\bf 1}&1&-N(F-1)&{6(F+1) \over F} \cr 
x_i^A &\Box&\Box&{\bf 1}&{\bf 1}&{-F \over F+5}&0&-{F+1 \over F+5} \cr 
\bar{C}_{li} &{\bf 1}&\Box&{\bf 1}&~\Box^{\ast}&{F \over N+F+4}&0&1 \cr 
\bar{S}_{AB} &~\Box\;\!\!\!\Box^{\ast}&{\bf 1}&{\bf 1}&{\bf 1}&{2F \over F+5}
&0&{4(F+3) \over F+5} \cr 
H^{lm} &{\bf 1}&{\bf 1}&{\bf 1}&\Box\;\!\!\!\Box&{-2F \over N+F+4}&0&0 }
\end{equation}
with the superpotential 
\begin{equation}
W=Mx\bar{C}\bar{q}+Np\bar{q}+\bar{S}p^2u+\bar{S}x^2+H\bar{C}^2.
\label{superpotential;dual2}
\end{equation}
We refer to this as the dual II theory. 
One can see that the 't Hooft anomaly matching conditions are satisfied. 

Notice that we found the duality between the expanded and the dual II theory 
only for $\Lambda_{SO(N+5)} \ll \Lambda_{SU(N)}$. Assuming holomorphy for 
$\Lambda_{SU(N)}/\Lambda_{SO(N+5)}$, 
however, the duality is valid for all $\Lambda_{SU(N)}/\Lambda_{SO(N+5)}$ 
(see also \cite{Prod}). 
Noting that the expanded theory (\ref{expanded;theory}) is equivalent to 
the electric theory (\ref{electric;theory}) for 
$\Lambda_{SU(N)} \ll \Lambda_{SO(N+5)}$, the dual II theory 
(\ref{dual2;theory}) turns out to be the dual of the electric theory. 

It is found that the gauge invariant operators (\ref{gaugeinvariants}) are 
mapped to the dual II theory in the following way 
\begin{eqnarray}
U &\rightarrow& \epsilon^{A_1 \cdots A_{F+5}}\epsilon^{B_1 \cdots B_{F+5}}
\epsilon^{x_1 \cdots x_F}\epsilon^{y_1 \cdots y_F}
\bar{q}_{x_1A_1}\cdots \bar{q}_{x_FA_F}
\bar{q}_{y_1B_1}\cdots \bar{q}_{y_FB_F} \nonumber \\
&&\times\bar{S}_{A_{F+1}B_{F+1}}\cdots\bar{S}_{A_{F+5}B_{F+5}}, \nonumber \\
\bar{B}^{~[l_1 \cdots l_N]} &\rightarrow& \epsilon_{A_1 \cdots A_{F+5}}
\left(x_{i_1}^{A_1}\bar{C}_{l_1i_1}\right) \cdots
\left(x_{i_{F+4}}^{A_{F+4}}\bar{C}_{l_{F+4}i_{F+4}}\right)p^{A_{F+5}}, 
\nonumber \\
V_k^{~[x_1 \cdots x_k][l_1 \cdots l_{N-k}]} &\rightarrow&
\tilde{b}^{~[A_1 \cdots A_{F-k}]}_{~[l_1 \cdots l_{F+4+k}]}
\bar{q}_{x_1A_1} \cdots \bar{q}_{x_{F-k}A_{F-k}},
\label{mapping}
\end{eqnarray}
where $\tilde{b}$ is the $SO(2F+8)$-invariant operator given by 
\begin{equation}
\tilde{b}^{~[A_1 \cdots A_{F-k}]}_{~[l_1 \cdots l_{F+4+k}]}=
\epsilon_{i_1 \cdots i_{2F+8}}x_{i_1}^{A_1}\cdots x_{i_{F-k}}^{A_{F-k}}
\bar{C}_{l_1i_{F+1-k}}\cdots \bar{C}_{l_{k+F+4}i_{2F+4}}
W_{i_{2F+5}i_{2F+6}}W_{i_{2F+7}i_{2F+8}}.
\end{equation}
$W$ is the $SO(2F+8)$ field strength superfield. 

We have derived the dual theory (\ref{dual2;theory}) 
by using the technique of deconfining and the Seiberg's duality naively. 
We can, however, obtain another dual theory by dualizing 
(\ref{dual2;theory}) again. We first think of the dual II theory 
(\ref{dual2;theory}) as $SU(F+5)$ gauge theory with a symmetric tensor. 
Then it can be dualized following the procedures we have discussed. 
The gauge group of the dual theory is 
$SU(F+5) \times SO(2F+8)_1 \times SO(2F+8)_2$, where $SO(2F+8)_1$ is the 
gauge group of the dual II theory and $SO(2F+8)_2$ 
results from dualizing the $SU(F+5)$. 
The vector index of the $SO(2F+8)_2$ is denoted by $i^{\prime}$. 
It is easy to see that the dual of the dual II theory has 
the following matter contents: 
\begin{equation}
\label{dual;dual2}
\matrix{ &SU(F+5)&SO(2F+8)_1&SO(2F+8)_2&SU(F)&SU(N+F+4) \cr 
u &{\bf 1}&{\bf 1}&{\bf 1}&{\bf 1}&{\bf 1} \cr 
M^{xl} &{\bf 1}&{\bf 1}&{\bf 1}&\Box&\Box \cr 
N^x &{\bf 1}&{\bf 1}&{\bf 1}&\Box&{\bf 1} \cr 
\bar{C}_{li} &{\bf 1}&\Box&{\bf 1}&{\bf 1}&~\Box^{\ast} \cr 
H^{lm} &{\bf 1}&{\bf 1}&{\bf 1}&{\bf 1}&\Box\;\!\!\!\Box \cr
\tilde{M}_x &{\bf 1}&{\bf 1}&{\bf 1}&\Box^{\ast}&{\bf 1} \cr
\tilde{M}_{xi} &{\bf 1}&\Box&{\bf 1}&\Box^{\ast}&{\bf 1} \cr
\tilde{H} &{\bf 1}&{\bf 1}&{\bf 1}&{\bf 1}&{\bf 1} \cr
\tilde{H}_i &{\bf 1}&\Box&{\bf 1}&{\bf 1}&{\bf 1} \cr
\tilde{H}_{ij} &{\bf 1}&\Box\;\!\!\!\Box&{\bf 1}&{\bf 1}&{\bf 1} \cr
\tilde{C}_{i^{\prime}} &{\bf 1}&{\bf 1}&\Box&{\bf 1}&{\bf 1} \cr
\tilde{C}_{ii^{\prime}} &{\bf 1}&\Box&\Box&{\bf 1}&{\bf 1} \cr
\tilde{S}^{AB} &\Box\;\!\!\!\Box&{\bf 1}&{\bf 1}&{\bf 1}&{\bf 1} \cr
\tilde{x}_{i^{\prime}A} &\Box^{\ast}&{\bf 1}&\Box&{\bf 1}&{\bf 1} \cr
\tilde{p}_A &\Box^{\ast}&{\bf 1}&{\bf 1}&{\bf 1}&{\bf 1} \cr
\tilde{u} &{\bf 1}&{\bf 1}&{\bf 1}&{\bf 1}&{\bf 1} \cr
\tilde{N}_{x} &{\bf 1}&{\bf 1}&{\bf 1}&\Box^{\ast}&{\bf 1} \cr
\tilde{q}^{xA} &\Box&{\bf 1}&{\bf 1}&\Box&{\bf 1} \cr
}
\end{equation}
with the superpotential 
\begin{eqnarray}
&&W=M^{xl}\bar{C}_{li}\tilde{M}_{xi}+N^x\tilde{M}_x+\tilde{H}u+
{\rm tr} ( \tilde{H}_{ij} )
+H^{lm}\bar{C}_{li}\bar{C}_{mi}  \nonumber \\
&&+\tilde{M}_x\tilde{x}_{i^{\prime}A}\tilde{C}_{i^{\prime}}\tilde{q}^{xA}+
\tilde{M}_{xi}\tilde{x}_{i^{\prime}A}\tilde{C}_{ii^{\prime}}\tilde{q}^{xA}+
\tilde{N}_x\tilde{p}_A\tilde{q}^{xA}+
\tilde{S}^{AB}\tilde{p}_A\tilde{p}_B\tilde{u}+
\tilde{S}^{AB}\tilde{x}_{i^{\prime}A}\tilde{x}_{i^{\prime}B} \nonumber \\
&&+\tilde{H}\tilde{C}_{i^{\prime}}\tilde{C}_{i^{\prime}}+ 
\tilde{H}_i\tilde{C}_{ii^{\prime}}\tilde{C}_{i^{\prime}}+
\tilde{H}_{ij}\tilde{C}_{ii^{\prime}}\tilde{C}_{ji^{\prime}}.
\label{W;1st}
\end{eqnarray}
Here we define the gauge invariant fields as follows 
\begin{eqnarray}
&&\tilde{M}_x=p^A\bar{q}_{xA}, \quad \tilde{M}_{xi}=x_i^A\bar{q}_{xA}, 
\nonumber \\
&&\tilde{H}=\bar{S}_{AB}p^Ap^B, \quad \tilde{H}_i=\bar{S}_{AB}p^Ax_i^A, 
\quad \tilde{H}_{ij}=\bar{S}_{AB}x_i^Ax_i^B. 
\end{eqnarray}
From (\ref{W;1st}), $N^x, \tilde{M}_x, \tilde{H}$ and $u$ become massive, and 
hence can be integrated out. 
Using the equation of motion for $\tilde{H}_{ij}$, 
we find that the field $\tilde{C}$ get a vev, so that the dual theory 
(\ref{dual;dual2}) is higgsed to $SU(F+5) \times SO(2F+8)$, where $SO(2F+8)$ 
is the diagonal subgroup of $SO(2F+8)_1 \times SO(2F+8)_2$. The vev makes 
$\tilde{H}_{ij}, \tilde{H}_i,$ and $\tilde{C}_{i^{\prime}}$ massive, and thus 
they can be integrated out. As a consequence, (\ref{dual;dual2}) becomes 
\begin{equation}
\label{dual;dual2.1}
\matrix{ &SU(F+5)&SO(2F+8)&SU(F)&SU(N+F+4) \cr 
M^{xl} &{\bf 1}&{\bf 1}&\Box&\Box \cr 
\bar{C}_{li} &{\bf 1}&\Box&{\bf 1}&~\Box^{\ast} \cr 
H^{lm} &{\bf 1}&{\bf 1}&{\bf 1}&\Box\;\!\!\!\Box \cr
\tilde{M}_{xi} &{\bf 1}&\Box&\Box^{\ast}&{\bf 1} \cr
\tilde{S}^{AB} &\Box\;\!\!\!\Box&{\bf 1}&{\bf 1}&{\bf 1} \cr
\tilde{x}_{iA} &\Box^{\ast}&\Box&{\bf 1}&{\bf 1} \cr
\tilde{p}_A &\Box^{\ast}&{\bf 1}&{\bf 1}&{\bf 1} \cr
\tilde{u} &{\bf 1}&{\bf 1}&{\bf 1}&{\bf 1} \cr
\tilde{N}_{x} &{\bf 1}&{\bf 1}&\Box^{\ast}&{\bf 1} \cr
\tilde{q}^{xA} &\Box&{\bf 1}&\Box&{\bf 1} \cr
}
\end{equation}
with the superpotential 
\begin{eqnarray}
W=M^{xl}\bar{C}_{li}\tilde{M}_{xi}+
\tilde{M}_{xi}\tilde{x}_{iA}\tilde{q}^{xA}+ 
\tilde{N}_x\tilde{p}_A\tilde{q}^{xA}+
\tilde{S}^{AB}\tilde{p}_{A}\tilde{p}_{B}\tilde{u}+
\tilde{S}^{AB}\tilde{x}_{iA}\tilde{x}_{iB}
+H^{lm}\bar{C}_{li}\bar{C}_{mi},
\label{W;2nd}
\end{eqnarray}
where $i$ is the vector index of the diagonal subgroup $SO(2F+8)$. 
Note that only in the case $F=0$, 
the dual (\ref{dual;dual2.1}) is exactly the same 
as the dual II theory (\ref{dual2;theory}).

In order to check the duality, turn on a vacuum expectation value in the 
electric theory. 
The moduli space of the 
electric theory is determined by the $D$-term conditions 
\begin{equation}
Q^{\ast}_{xA}Q^{xB}-\bar{Q}^l_A\bar{Q}_l^{\ast B}+S_{AC}^{\ast}S^{CB}
=c\delta^B_A,
\label{d-term}
\end{equation}
where $c$ is a constant. 
We turn on vev's for $Q$ and $\bar{Q}$ in the way 
\begin{equation}
Q^{xA}=0, \quad \bar{Q}^{l}_{A}=
\left( \begin{array}{@{\,}ccc|@{\,}cc}
\bar{a}_1 &        &           &&   \\ 
          & \ddots &           &&  \\ 
          &        & \bar{a}_r &&   \\ \hline 
          &        &           &&   \\
          &       &            &&  \\
          &        &           &&
\end{array} 
\right).
\end{equation}
The vev's of $S^{AB}$ are fixed by the $D$-term conditions. 
When $c=0$, 
the electric theory is higgsed to $SU(N-r)$ and its massless field 
contents are the $F$ fundamentals $Q^{x\hat{A}}$, the $N+F+4-r$ anti-fundamentals 
$\bar{Q}^{\hat{l}}_{\hat{A}}$, and a symmetric tensor $S^{\hat{A}\hat{B}}$, 
where $\hat{A}=r+1, \cdots, N,$ and $\hat{l}=r+1, \cdots, N+F+4.$ 
Note also that $H^{lm}$  gets a vev 
of rank $r$ and the other gauge invariant operators take vanishing values. 
Turning to the dual II theory, one can see from the superpotential 
(\ref{superpotential;dual2}) that 
$\bar{C}_{li}, \quad l=1,\cdots, r$ become massive because $H^{lm}$ 
has a vev of rank $r$. Integrating out the massive fields, the symmetries 
of the dual II theory are broken as follows 
\begin{eqnarray}
&&SU(F+5) \times SO(2F+8) \times SU(F) \times SU(N+F+4) \nonumber \\
&&\rightarrow SU(F+5) \times SO(2F+8) \times SU(F) \times SU(N+F+4-r).
\end{eqnarray}
This is indeed the dual of the electric theory. 

Let us now consider the case $c \ne 0$. The electric theory is higgsed to 
$SO(N-r)$, and its contents of massless fields are $N+2F+4-r$ vectors. 
As in the case $c=0$, $H^{lm}$ has a vev of rank $r$. One of the most 
important differences is that 
the gauge invariant operator $U=\det S$ takes a non-vanishing value. 
Turning to the dual II theory, it follows from (\ref{mapping}) that the 
fields $\bar{q}_{xA}$ and $\bar{S}_{AB}$ get vev's 
so that the gauge 
group $SU(F+5)$ is higgsed to $SO(5)$ while $SO(2F+8)$ is unbroken. 
Up to gauge and flavor transformations, the vev's are given by 
\begin{equation}
\bar{q}_{xA}=
\left( \begin{array}{@{\,}ccc|@{\,}c}
 b      &         &     &    \\
        & \ddots  &     & ~00000\! \\
        &         & b   &       
\end{array} 
\right), \quad 
\bar{S}_{AB}=
\left( \begin{array}{@{\,}ccc|@{\,}ccc}
&& &   &     &         \\ 
&{\bf{0}}_{F \times F}&&&& \\
&&&&& \\ \hline
&&     &   &     &    \\ 
&&     &   & b{\bf 1}_{5 \times 5} &       \\ 
&&     &         &        &    
\end{array} 
\right),
\label{vev;dual2}
\end{equation}
where $b$ is written in terms of $\bar{a}_i$ and $c$. 
Substituting them into 
the superpotential (\ref{superpotential;dual2}) and integrating out the massive 
fields, the dual II theory reduces to the supersymmetric $SO(5) \times SO(2F+8)$ 
gauge theory with the following matter contents 
\begin{equation}
\label{dual2;flat}
\matrix{ &SO(5)&SO(2F+8)&SU(N+2F+4-r) \cr 
\bar{q}_{\mu i} &{\bf 1}&\Box&\Box^{\ast} \cr 
M^{\mu\nu} &{\bf 1}&{\bf 1}&\Box\;\!\!\!\Box \cr 
p^{A} &\Box&{\bf 1}&{\bf 1} \cr
u &{\bf 1}&{\bf 1}&{\bf 1}
}
\end{equation}
with the superpotential 
\begin{equation}
W=p^Ap^Au + M^{\mu\nu}q_{\mu i}q_{\nu i},
\end{equation}
where $SO(2F+8)$ vector fields $q_{\mu i},~ \mu =1,2,\cdots, N+2F+4-r$ 
and the meson fields $M^{\mu\nu}=M^{\nu\mu}$ are given by 
\begin{equation}
q_{\mu i}=(x_i^{A^{\prime}} , \bar{C}_{\hat{l}i}), \quad 
M^{\mu\nu}=
\left( \begin{array}{@{\,}c|c@{\,}}
\bar{S}_{A^{\prime}B^{\prime}} & M^{x\hat{l}} \\ \hline 
\left(M^{x\hat{l}}\right)^T & H^{\hat{l}\hat{m}} 
\end{array} 
\right), 
\end{equation}
where $A^{\prime},B^{\prime}=1,2,\cdots, F$. 
Note that (\ref{dual2;flat}) can be regarded as supersymmetric $SO(5)$ 
gauge theory with one flavor. It follows from \cite{IS;so} that the exact 
superpotential of (\ref{dual2;flat}) is given by 
\begin{equation}
W = {1 \over 2}(\epsilon_L + \epsilon_R){4\Lambda^4 \over \sqrt{M}} + Mu 
+ M^{\mu\nu}\bar{q}_{\mu i}\bar{q}_{\nu i}. 
\end{equation}
Here $\epsilon_{L, R}=\pm 1, ~M=p^Ap^A$ and $\Lambda$ is the dynamical scale of 
the $SO(5)$ SUSY gauge theory. 
Depending on $\epsilon_L \epsilon_R = \pm 1$, the moduli space 
consists of two branches. For $\epsilon_L \epsilon_R = 1$, it is obvious that 
no SUSY vacuum exists. On the other hand, for $\epsilon_L \epsilon_R = -1$,  
no dynamical superpotential can be generated so that $M$ and $u$ become 
massive. 
Thus the relation between the deformed electric and dual II theories is indeed 
the duality found by Seiberg \cite{Sei2}.

Finally we study the duality by adding 
$\lambda_{lm}S^{AB}\bar{Q}^l_A \bar{Q}^m_B, \quad \lambda=\lambda^T$, 
to the superpotential. For convenience, we assume that $\lambda_{lm}$ takes the 
following form 
\begin{equation}
\lambda_{lm}
= {\rm diag}( \underbrace{\lambda, \cdots, \lambda}_{k}, 
\underbrace{0, \cdots, 0}_{N+F+4-k} ).
\end{equation}
In the electric theory, there exists a 
branch of the moduli space parameterized by 
\begin{equation}
Q=\bar{Q}=0, \quad S^{AB} \propto \delta^{AB}. 
\end{equation}
By deforming the electric theory along the flat direction, the electric 
theory flows to $SO(N)$ SUSY gauge theory with $N+2F+4-k$ flavors. 
On the other hand, the superpotential of the dual II theory is given by 
\begin{equation}
W = Mx\bar{C}\bar{q} + Np\bar{q} + \bar{S}p^2u + \bar{S}x^2 + H\bar{C}^2 
+ \lambda\sum_{l=1}^{k}H^{ll}. 
\label{superpotential;deformed}
\end{equation}
Recall that we are considering the branch $U \ne 0$. This means that $\bar{q}$ 
and $\bar{S}$ take the same form of vev's as (\ref{vev;dual2}). Using the 
$F$-term condition for $H^{lm}$, $\bar{C}$ is found to take a vev which breaks 
the gauge group $SO(2F+8)$ to $SO(2F+8-k)$. Expanding the dual II theory 
around the vacuum and integrating out massive fields, it is found that 
the dual II theory reduces to: 
\begin{equation}
\label{dual2;Yukawa}
\matrix{ &SO(5)&SO(2F+8-k)&SU(N+2F+4-k) \cr 
\bar{q}_{\mu \hat{i}} &{\bf 1}&\Box&\Box^{\ast} \cr 
M^{\mu\nu} &{\bf 1}&{\bf 1}&\Box\;\!\!\!\Box \cr 
p^{A} &\Box&{\bf 1}&{\bf 1} \cr
u &{\bf 1}&{\bf 1}&{\bf 1}
}
\end{equation}
with the superpotential 
\begin{equation}
W = p^Ap^Au + M^{\mu\nu}\bar{q}_{\mu \hat{i}}\bar{q}_{\nu \hat{i}}.
\end{equation}
Here we denote the vector index of the $SO(2F+8-k)$ by $\hat{i}$. 
From the same discussions as those in (\ref{dual2;flat}), we can see that 
(\ref{dual2;Yukawa}) flows to the dual of the deformed electric theory.

\vspace{1cm}

It is a pleasure to thank Prof. N. Sakai for a useful discussion. 
The author also thanks Christian Baraldo for careful reading of the manuscript.

%%%%%%%%%%%%%%%%%%%%%%%%%%%%%%%%%%%%%%%%%%%%%%%%%%%%%%%%%%%%%%%%%%%%%
%%%%%%%%%%%%%%%%%%%% Referrence %%%%%%%%%%%%%%%%%%%%%%%%%%%%%%%%%%%%%
%%%%%%%%%%%%%%%%%%%%%%%%%%%%%%%%%%%%%%%%%%%%%%%%%%%%%%%%%%%%%%%%%%%%%

%
%%%%%%%%%%%% figure caption %%%%%%%%%%%%%%%%%
%\def\epsfbox{}  %NO FIGS!!!
%
%\newpage
%\section*{Figure captions}
%\begin{itemize}
%\item[Fig.\ 1]
%Effective potential for the (A, P) case with $M=1$ .
%
%\item[Fig.\ 2]
%Effective potential for the (A, A) case with $M=1$ .
%
%\item[Fig.\ 3]
%Effective potential for the (P, A) case.
%
%
%\end{itemize}

%
%%%%%%%%%% figure %%%%%%%%%%%%%%%%%%%%%%%%%%
%
%
%
%\newpage
%\begin{figure}
% \leavevmode
% \epsfysize=10cm
% \centerline{\epsfbox{APV.eps}}
% \caption{
%Effective potential for the (A, P) case with $M=1$ .
%}
% \label{fig:APV}
%\end{figure}


\newpage
\begin{thebibliography}{100}


\bibitem{Sei1} N. Seiberg, {\it Phys.\ Rev.\ }{\bf D49} (1994) 6857, 
{\tt hep-th/9402044}.

\bibitem{IRSei} K. Intriligator, R. G. Leigh and N. Seiberg, 
{\it Phys.\ Rev.\ }{\bf D50} (1994) 1092, {\tt hep-th/9403198};
K. Intriligator, {\it Phys.\ Lett.\ }{\bf B336} (1994) 409,
{\tt hep-th/9407106}.

\bibitem{IS;lec} K. Intriligator and N. Seiberg, 
{\it Lectures on Supersymmetric Gauge Theories and Electric-Magnetic Duality},
{\tt hep-th/9509066}.

\bibitem{Sei2} N. Seiberg, {\it Nucl.\ Phys.\ }{\bf B435} (1995) 129, 
{\tt hep-th/9411149}.

\bibitem{KSS} D. Kutasov, {\it Phys.\ Lett.\ }{\bf B351} (1995) 230, 
{\tt hep-th/9503086}; 
D. Kutasov and A. Schwimmer, {\it Phys.\ Lett.\ }{\bf B354} (1995) 315, 
{\tt hep-th/9505004}; 
D. Kutasov, A. Schwimmer and N. Seiberg, {\it Nucl.\ Phys.\ }{\bf B459} 
(1996) 445, {\tt hep-th/9510222}. 

\bibitem{IS;so} K. Intriligator and N. Seiberg, 
{\it Nucl.\ Phys.\ }{\bf B444} (1995) 125, {\tt hep-th/9503179}.

\bibitem{IP;sp} K. Intriligator and P. Pouliot, 
{\it Phys.\ Lett.\ }{\bf B353} (1995) 471, {\tt hep-th/9505005}. 

\bibitem{IRSt} 
K. Intriligator, 
{\it Nucl.\ Phys.\ }{\bf B448} (1995) 187, {\tt hep-th/9505051}; 
R. G. Leigh and M. J. Strassler, 
{\it Phys.\ Lett.\ }{\bf B356} (1995) 492, {\tt hep-th/9505088};
K. Intriligator, R. G. Leigh and M. J. Strassler, 
{\it Nucl.\ Phys.\ }{\bf B456} (1995) 567, {\tt hep-th/9506148}.


\bibitem{Chi-nonChi} P. Pouliot, {\it Phys.\ Lett.\ }{\bf B359} (1995) 108, 
{\tt hep-th/9507018}; 
P. Pouliot and M. J. Strassler, {\it Phys.\ Lett.\ }{\bf B370} (1996) 76, 
{\tt hep-th/9510228}; 
P. Pouliot and M. J. Strassler, {\it Phys.\ Lett.\ }{\bf B375} (1996) 175, 
{\tt hep-th/9602031}; 
T. Kawano, {\it Prog.\ Theor.\ Phys.\ }{\bf 95} (1996) 963, 
{\tt hep-th/9602035}. 

\bibitem{Decon} 
M. Berkooz, {\it Nucl.\ Phys.\ }{\bf B452} (1995) 513, {\tt hep-th/9505067}; 
P. Pouliot, {\it Phys.\ Lett.\ }{\bf B367} (1996) 151, {\tt hep-th/9510148}; 
M. A. Luty, M. Schmaltz and J. Terning, 
{\it Phys.\ Rev.\ }{\bf D54} (1996) 7815, {\tt hep-th/9603034}. 

\bibitem{Prod} E. Poppitz, Y. Shadmi and S.P. Trievi, 
{\it Nucl.\ Phys.\ }{\bf B480} (1996) 125, {\tt hep-th/9605113}, 
{\it Phys.\ Lett.\ }{\bf B388} (1996) 561, {\tt hep-th/9606184}. 



\bibitem{Excep} I. Pesando, {\it Mod.\ Phys.\ Lett.\ }{\bf A10} (1995) 1871, 
{\tt hep-th/9506139}; 
S. B. Giddings and J. M. Pierre, {\it Phys.\ Rev.\ }{\bf D52} (1995) 6065, 
{\tt hep-th/9509196}; 
P. Ramond, {\tt hep-th/9608077}; 
J. Distler and A. Karch, {\tt hep-th/9611088}. 


\bibitem{Aha} 
O. Aharony, {\it Phys.\ Lett.\ }{\bf B351} (1995) 220, {\tt hep-th/9502013}; 
O. Aharony, J. Sonnenschein and S. Yankielowicz, 
{\it Nucl.\ Phys.\ }{\bf B449} (1995) 509, {\tt hep-th/9504113}. 

\bibitem{RS}
R. G. Leigh and M. J. Strassler, 
{\it Nucl.\ Phys.\ }{\bf B447} (1995) 95, {\tt hep-th/9503121}.


\bibitem{Maru}
N. Maru and S. Kitakado, {\tt hep-th/9609230}; 
J. H. Brodie and M. J. Strassler, {\tt hep-th/9611197}. 



\end{thebibliography}
\end{document}